\documentclass[aps,prd,secnumarabic,amssymb,nofootinbib,amsmath,11pt]{revtex4} %twocolumn
\usepackage{amsfonts,amsmath,hyperref,url, color}
\usepackage{bm, bbm}
\usepackage{graphicx}
\usepackage{mathtools}
\usepackage{nccmath}
\usepackage[utf8]{inputenc}
\usepackage[english]{babel}
\usepackage{amsmath,amsfonts,amssymb}
\usepackage{bbold}
\usepackage{hyperref}
\usepackage{caption,adjustbox}
\usepackage{multirow}
\usepackage{cleveref}
\usepackage{fontawesome}
\usepackage{comment}
\usepackage{graphicx}
\usepackage{bbding}
\usepackage[usenames,dvipsnames]{xcolor}
\usepackage[arrowdel]{physics}
\usepackage{tensor}
\usepackage{cancel}
\usepackage{tikz}
\usepackage{listings} 
\usepackage[title,toc,titletoc]{appendix}
\usepackage{titlesec}
\usepackage{tocloft}

\usepackage{color}

\newcommand{\be}{\begin{equation}}
\newcommand{\ben}{\begin{equation*}}
\newcommand{\ee}{\end{equation}}
\newcommand{\een}{\end{equation*}}
\newcommand{\ba}{\begin{eqnarray}}
\newcommand{\ea}{\end{eqnarray}}

\begin{document}

\title{Entanglement entropy in conformal quantum mechanics}

\author{Michele Arzano}
\email{michele.arzano@na.infn.it}
\affiliation{Dipartimento di Fisica ``E. Pancini", Universit\`a di Napoli Federico II, I-80125 Napoli, Italy\\}
\affiliation{INFN, Sezione di Napoli,\\ Complesso Universitario di Monte S. Angelo,\\
Via Cintia Edificio 6, 80126 Napoli, Italy}

\author{Alessandra D'Alise}
\email{alessandra.dalise@unina.it}
\affiliation{Dipartimento di Fisica ``E. Pancini", Universit\`a di Napoli Federico II, I-80125 Napoli, Italy\\}
\affiliation{INFN, Sezione di Napoli,\\ Complesso Universitario di Monte S. Angelo,\\
Via Cintia Edificio 6, 80126 Napoli, Italy}

\author{Domenico Frattulillo}
\email{domenico.frattulillo@unina.it}
\affiliation{Dipartimento di Fisica ``E. Pancini", Universit\`a di Napoli Federico II, I-80125 Napoli, Italy\\}
\affiliation{INFN, Sezione di Napoli,\\ Complesso Universitario di Monte S. Angelo,\\
Via Cintia Edificio 6, 80126 Napoli, Italy}

\begin{abstract}
We consider sets of states in conformal quantum mechanics associated to generators of time evolution whose orbits cover different regions of the time domain. States labelled by a continuous global time variable define the two-point correlation functions of the theory seen as a one-dimensional conformal field theory. Such states exhibit the structure of a thermofield double built on bipartite eigenstates of generators of non-global time evolution. In terms of the correspondence between radial conformal symmetries in Minkowski spacetime and time evolution in conformal quantum mechanics proposed in \cite{Arzano:2020thh,Arzano:2021cjm} such generators coincide with conformal Killing vectors tangent to  worldlines of Milne and diamond observers at constant radius. The temperature of the thermofield double states in conformal quantum mechanics reproduces the temperatures perceived by such diamond and Milne observers. We calculate the entanglement entropy associated to the thermofield double states and obtain a UV divergent logarithmic behaviour analogous to known results in two-dimensional conformal field theory in which the entangling boundary is point-like.
\end{abstract}

\maketitle

\section{Introduction}

Entanglement entropy plays a central role in our understanding of the interplay between the quantum realm and the geometry of spacetime. In quantum field theory, it can characterize the quantum correlations between degrees of freedom inside and outside a given region of spacetime. Due to the presence of short scale correlations such geometric entropy is generally UV divergent but has the notable feature that the leading divergent term is proportional to the area of the entangling surface. Such property suggests a link between entanglement entropy and thermodynamic properties of black holes. Indeed, according to the celebrated entropy-area relation \cite{Bekenstein:1973ur}, black holes are characterized by an entropy proportional to their horizon surface area. While there is to date no general consensus on the origin of the degrees of freedom responsible for such entropy \cite{Jacobson:1999mi,Carlip:2007ph}, its area scaling suggests that black hole entropy is intimately related to quantum correlations across the horizon \cite{Solodukhin:2011gn}.

Entanglement entropy is notoriously hard to compute in quantum field theory (see \cite{Rangamani:2016dms} for a review of the various techniques). One notable exception where analytical results exist is two-dimensional conformal field theory (CFT$_2$) \cite{Holzhey:1994we,Calabrese:2004eu}. In this letter we show that $0+1$-dimensional conformal field theory, i.e. conformal quantum mechanics, provides yet another model in  which entanglement entropy associated to correlations of the two-point function can be calculated analytically in a rather straightforward way.

Our analysis is motivated by the correspondence between radial conformal Killing vectors in Minkowski space-time and the generators of conformal transformations of the real line which describe alternative time evolutions in conformal quantum mechanics. In \cite{Arzano:2020thh,Arzano:2021cjm} it was shown that such alternative time evolutions coincide with the tangent vectors to worldlines of observers sitting at the origin within a causal diamond and in Milne space-time.
The definition of positive and negative frequency modes for a conformally invariant field for such observers are different than the one of inertial observers and lead to the construction of different Fock spaces with different vacuum states \cite{Olson:2010jy,Su:2015oys,Higuchi:2017gcd,Wald:2019ygd}. 
The inertial vacuum density matrix is an entangled state between the modes associated to such observers and the ones defined in the space-time region from which they cannot exchange signals with. Tracing over the inaccessible degrees of freedom the vacuum density matrix becomes a thermal state at a characteristic diamond or Milne temperature. In conformal quantum mechanics one can identify states which have a similar structure associated to time domains covered by orbits of different generators of time evolution and find analogous temperatures \cite{Arzano:2020thh,Arzano:2021cjm}. In particular states labelled by a continuous time variable are used to build a two-point function which corresponds the restriction to wordlines of observers sitting at the origin of the two-point function of a massless scalar field in Minkwoski space-time defined on the inertial vacuum state. Such states exhibit the structure of a thermofield double in terms of excitations of the Hamiltonian which generates time translations in time domains with boundaries.
Here we show how it is possible to quantify the entanglement of such state in terms of the Von Neumann entropy of the associated reduced density matrix obtained by tracing over one set of degrees of freedom of the thermofield double.
The result diverges logarithmically when the UV regulator is sent to zero as in two-dimensional conformal field theory, a result expected in models where the entangling boundary is point-like. 

In the next section we start by recalling the correspondence between radial conformal Killing vectors in 3+1-dimensional Minkowski space time and conformal transformations of the real line with particular focus on generators whose orbits exhibit boundaries. In Section 3 we examine the role of such generators in conformal quantum mechanics, we introduce the sets of states naturally associated to them and spell out their significance in the definition of the two-point function of the model seen as a one-dimensional conformal field theory. In Section 4 we show how states labelled by a continuous time variable, whose inner product gives the two-point function of the theory, exhibit the structure of a thermofield double state in terms of the excitations of Hamiltonians whose orbits exhibit boundaries on the time domain. We point out how, after tracing over one set of degrees of freedom of the thermofield double, one obtains thermal density matrices at the Milne and diamond temperature. In Section 5 we proceed to evaluate the entanglement entropy of the reduced vacuum density matrix. This requires an appropriate regularization of the vacuum state in order to control its non-normalizability which is directly related to the divergence of the two-point function at coincident points. We conclude in Section 6 with a summary and comments on the results obtained.

\section{Radial conformal Killing vectors in Minkowski spacetime and conformal transformations of the real line}

Let us consider the Minkowski metric in spherical coordinates 
\be
d s^2 = -d t^2 + d r^2 + r^2 (d \theta^2 + \sin^2 \theta\,  d\phi^2)\,.
\ee
A vector field $\xi$ is a conformal Killing vector if
\be
\mathcal{L}_{\xi} g_{\mu\nu} \propto g_{\mu\nu}
\ee
where $g_{\mu\nu}$ is a generic metric and $\mathcal{L}_{\xi}$ denotes the Lie derivative. For Minkowski spacetime all {\it radial} conformal Killing vectors were classified in \cite{RCM} and they have the general form
\be\label{xi11}
\xi = \left(a(t^2+r^2)+b t +c\right)\, \partial_t + r (2 a t + b)\, \partial_r
\ee
with $a,b,c$ real constants. Central to the present work is the fact that such vectors can be written as 
\be
\xi = a K_0 + b D_0 + c P_0\,,
\ee
with $P_0$, $D_0$ and $K_0$ generating, respectively, time translations, dilations and special conformal transformations
\be
P_0  =  \partial_t\,,\qquad D_0  =  r\, \partial_r + t\, \partial_t\,,\qquad K_0  =  2 t r\, \partial_r + (t^2+r^2)\, \partial_t \,,
\ee
whose commutators close the $\mathfrak{sl}(2,\mathbb{R})$ Lie algebra 
\be
[P_0,D_0]= P_0\,,\qquad [K_0,D_0]= - K_0\,,\qquad [P_0,K_0]= 2 D_0\,.
\ee

The causal character of different conformal Killing vectors changes according to the values of the real constants $a,b$ and $c$ \cite{RCM}. For example, when $a=0$ and $b = 0$ the conformal Killing vectors are everywhere time-like. This is the case of the vector $P_0$ which generates evolution in {\it inertial time} and whose integral lines are worldlines of static inertial observers i.e. straight infinite lines with $r=\mathrm{const}$. In general, however, such conformal Killing vectors are not everywhere time-like. For example generators for which $a=0$ and $b\neq 0$ are null on the light-cone emanating from the point $(t= -c/b,r=0)$, time-like inside such light-cone and space-like outside.  The generator of dilations $D_0$ is one of such vector whose integral lines are straight lines emanating from the origin in the $t$-$r$ plane and within the future oriented light-cone describe worldlines of comoving observers in a expanding Milne universe \cite{Ling:2018tih} (contracting in the past cone). The Milne universe is a flat FRLW space-time with scale factor linear in time
\be
d s^2 = -d \bar{t}^{\,2} + \bar{t}^{\,2}\left(d \chi^2 + \sinh \chi^2 d\Omega^2\right)\,.
\ee 
Such metric is simply the Minkowski space-time metric restricted to the future cone and rewritten in hyperbolic slicing coordinates $t = \bar{t}\, \cosh \chi$ and $r = \bar{t}\, \sinh \chi$. Worldlines of comoving Milne observers are straight lines with $\chi=\mathrm{const.}$. These observers move with radial Minkowskian velocity $r/t= \tanh \chi$ which approaches the value of 1 as $\chi \rightarrow \infty$ i.e. as the worldline of the comoving observer approaches the light-cone. Notice how the time evolution of Milne comoving observers is not {\it eternal}, albeit being still infinite, since integral lines of $D_0$ have a beginning (and an end for the past cone) at the origin $(t=0, r=0)$. Finally notice that the conformal Killing $D_0$ vector becomes null on the light-cone which thus represents a {\it conformal Killing horizon} \cite{DyerHonig1979}. \\
% $\Delta = b^2 - 4 a c\,$

Another conformal Killing vector which will be relevant for our discussion is the following combination of translations and special conformal transformations
\be\label{sckv}
S_0 = \frac{1}{2}\left(\alpha P_0-\frac{K_0}{\alpha}\right)\ .
\ee
Such conformal Killing vector exhibits a more articulated causal structure determined by the intersection of two light-cones emanating from the points $(t=\pm\alpha, r=0)$: it is null on these light-cones, time-like inside or outside both light-cones and space-like elsewhere. Thus conformal transformations generated by the vector map the intersection of the two light-cones into itself \cite{Jacobson:2015hqa}, such region is a {\it causal diamond} of radius $\alpha$. The integral lines of $S_0$ within the diamond are worldlines of accelerated observers \cite{RCM} with a {\it finite life-time} (since they are restricted to the causal diamond) and take the form
\begin{equation}
    t^2-\left(r-\alpha \omega\right)^2=\alpha^2\left(1-\omega^2\right)
\end{equation}
parametrized by a dimensionless parameter $\omega\neq 0$.
% {\bf (add equation of integral lines taken from Herrero and Morales)}.
Notice how the generators of boosts, which describe time evolution in Rindler space-time, are not included in the above classification since they are not radial (written in spherical coordinates they are not independent of angular variables).\\ 

The connection with conformal quantum mechanics stems from the observation that along $r= \mathrm{const}$ worldlines (but also on light-cones $u=t-r= \mathrm{const}$, $v=t+r= \mathrm{const}$) the conformal Killing vector \eqref{xi11} takes the form

\be\label{xikv}
\xi =\left( a\, t^2 + b\, t + c\, \right) \partial_t\,.
\ee
This is the general expression for a generator of {\it conformal transformations of the real (time) line}. In particular $P_0 = \partial_t$ generates translations in ``inertial time" $t$ covering the entire time line. This time variable can be identified with the proper time of static inertial observers in Minkowski space-time (observers with four-velocity parallel to the (conformal) Killing vector $P_0$). The dilation Killing vector $D_0 = t \partial_t $ generates translation in ``Milne time" $\tau$ defined by 
\be 
D_0= \alpha \partial_\tau \,.
\ee
The only worldline at fixed radius in the Milne universe is the one at $r=0$ for which $\tau$ corresponds to the conformal time. One can easily show that 
\be
t = \pm \alpha\, \exp(\frac{\tau}{\alpha})
\ee
and thus the Milne time coordinate covers {\it only half} (the regions $t>0$ or $t<0$) of the whole time domain. Finally the conformal Killing vector  $S_0 = \frac{1}{2 \alpha} \left(\alpha^2- t^2\right) \partial_{t} $ generates translation in ``diamond time" $\sigma$  
\be
S_0 =\alpha \partial_\sigma \,.
\ee
In Minkowski space-time this vector corresponds to the restriction to the worldline of a static diamond observer at $r=0$ of the conformal Killing vector \eqref{sckv}.
On the time line this time variable is related to inertial time by the transformation
\be
t = \alpha\, \tanh{\left(\frac{\sigma}{2\alpha}\right)}
\ee
and thus it covers only the segment $|t|<\alpha$ of the time domain. 

%two point functions describe quantum correlation in the vacuum so they know about entanglement 

\section{Time evolution in conformal quantum mechanics and the CFT$_1$ two-point function}

The quantum mechanical counterpart of the generator \eqref{xikv} is
\be
G=i \xi = a K + b D + c H\,,
\ee
where $K=iK_0$, $D=iD_0$ and $H=iP_0$, corresponds to the most general Hamiltonian of {\it conformal quantum mechanics}, a quantum mechanical model first studied in \cite{deAlfaro:1976vlx} characterized by an inverse square potential Lagrangian invariant under the one-dimensional group of conformal transformations of the time axis $\mathrm{SL}(2, \mathbb{R})$. Such model can be understood as a one-dimensional conformal field theory \cite{Chamon:2011xk, Jackiw:2012ur}, as we briefly recall below.

In \cite{deAlfaro:1976vlx} two sets of states were constructed: the eigenstates $|n\rangle$ of the elliptic operator $R$ which has discrete spectrum and are normalizable, and states labelled by inertial time $|t\rangle$ on which the operator generating inertial time translations $H$ acts as a derivative
\be
H |t\rangle = -i \dv{}{t} |t\rangle\,.
\ee
Such states are labelled by the continuous parameter $t$ and are non-normalizable as we will show below. The action of the remaining $\mathrm{SL}(2, \mathbb{R})$ generators is given by
\begin{align}\label{real1}
     D \ket{t} &= -i \left(t\dv{}{t}+r_0\right)\ket{t}\\
   K \ket{t} &= -i \left(t^2\dv{}{t}+2r_0 t\right)\ket{t}\,.
\end{align}
Introducing ladder operators
\begin{equation}\label{algebraR}
    L_0= R=\frac{1}{2}\left(\frac{K}{\alpha}+\alpha H\right)\qquad
    L_\pm= S\pm iD = \frac{1}{2}\left(\frac{K}{\alpha}-\alpha H\right)\pm i D\, , 
\end{equation} 
whose commutators are
\begin{equation}\label{so21alg}
    \comm{L_0}{L_\pm}=\pm L_\pm, \qquad \comm{L_-}{L_+}=2L_0\, , 
\end{equation}
we define sates $|n\rangle$ labelled by positive integers $n=0,1,...$ on which the action of the operators \eqref{algebraR} is given by
\begin{align}
    L_0\ket{n}&=(n+r_0)\ket{n}\\
    L_\pm\ket{n}&=\sqrt{((n+r_0)(n+r_0\pm 1))-r_0(r_0-1))}\ket{n\pm 1}\,,
\end{align}
with the orthonormality relation
\be
\langle n| n'\rangle = \delta_{nn'}\,.
\ee
The constant $r_0$, the eigenvalue of the ground state $|n=0\rangle$, is a positive real number \cite{Jackiw:2012ur} and it is related to the eigenvalue of the Casimir operator
\be
\mathcal{C} = R^2-S^2-D^2 = \frac{1}{2}\left(HK+KH\right)-D^2\,,
\ee
given by 
\be
\mathcal{C} \ket{n} = r_0(r_0-1) \ket{n}\,.
\ee
For $r_0$ integer and half integer with $r_0 \ge 1$ the set of states $\ket{n}$ provides an irreducible representation of the Lie algebra $\mathfrak{sl}(2,\mathbb{R})$ belonging to the so-called {\it discrete series} \cite{Sun:2021thf}.\\

In what follows we focus on the case $r_0=1$ as in \cite{Arzano:2021cjm,Arzano:2020thh} since this is the case in which the two-point function conformal quantum mechanics, seen as a one dimensional conformal field theory, is equivalent to the restriction of the two-point function of a massless scalar field along the worldline of observers sitting at $r=0$ in Minkowski space-time.\\

Wavefunctions representing the states $|n\rangle$ as functions of the inertial time coordinate $t$ can be obtained by considering the action of the $R$ operator on the $\langle t|$ states obtaining the differential equation 
\begin{equation}
    \mel{t}{R}{n}=\frac{i}{2}\left[\left(\alpha+\frac{t^2}{\alpha}\dv{}{t}\right)+2 \frac{t}{\alpha}\right]\ip{t}{n} =(n+1)\ip{t}{n}
\end{equation}
whose solution is given by
\begin{equation}\label{iptn}
  \ip{t}{n}= - \frac{\alpha^2c_n e^{2 i (n+1) \tan ^{-1}\left(\frac{\alpha}{t }\right)}}{\alpha^2+t ^2}\ .
\end{equation}
One can determine the normalization constants by iteration (see Appendix \ref{cn}) obtaining
\begin{equation}
    c_n=\sqrt{\frac{\Gamma(2+n)}{n!}}\ .
\end{equation} 
Plugging this expression in \eqref{iptn} we can obtain the following equation relating $|t\rangle$ and $|n\rangle$ states\footnote{Where we used the relation $2i\tan^{-1}{x}=\log{\frac{1+i x}{1-i x}}$ in \eqref{iptn} to write \eqref{tnov}.}
\begin{equation}
\label{tnov}
\ket{t}=\left(\frac{\frac{\alpha+it}{\alpha-it}+1}{2}\right)^{2}\ \sum_{n} (-1)^n \left(\frac{\alpha+it}{\alpha-it}\right)^n\ \sqrt{\frac{\Gamma(2+n)}{n!}}\ket{n}\ .
\end{equation}
These states can be written in terms of the action of the creation operator $L_+$ on the ground state $\ket{n=0}$ 
\begin{equation}
\label{tnzero}
\ket{t}=\left(\frac{\frac{\alpha+it}{\alpha-it}+1}{2}\right)^{2}\ \exp{-\left(\frac{\alpha+it}{\alpha-it}\right)\, L_+} \ket{n=0}\,,
\end{equation}
and in particular 
\begin{equation}
\label{tzeron}
\ket{t=0}= \exp\left(- L_+\right) \ket{n=0}\,.
\end{equation}

As discussed in \cite{Chamon:2011xk,Jackiw:2012ur} the inner product between the $|t\rangle$ states can be interpreted as the two-point function of a one-dimensional CFT. One can explicitly evaluate such two-point function using \eqref{tnov}
\begin{equation}\label{twpf}
    G(t_1,t_2)\equiv  \braket{t_1}{t_2}=-\frac{ \alpha^{2 }}{4\ (t_1-t_2)^{2 }}\,.
\end{equation}
Notice how such expression matches, modulo a constant factor, that of the two-point function of a free massless scalar field in Minkowski space-time along the trajectory of a static inertial observer sitting at $r=0$ \cite{Arzano:2020thh}. Moreover, since the Hamiltonian $H$ generates the time evolution, the $t$-state can actually be obtained with a time translation from the $t=0$ vacuum
\begin{equation}
    \ket{t}= e^{i H t}\ket{t=0}=\frac{1}{4} e^{(\alpha+it) H}\ket{n=0}
\end{equation}
and thus the two-point function \eqref{twpf} can be written as 
\be\label{2pfvac}
G(t_1,t_2) = \braket{t_1}{t_2} = \bra{t=0} e^{-i H t_1}\, e^{i H t_2}\ket{t=0}\,.
\ee
It is instructive to look at the two-point function \eqref{2pfvac} in terms of eigenstates of the generator $H$. These eigenstates $\ket{E}$, first introduced in \cite{deAlfaro:1976vlx}, are defined by
\be
H\ket{E}=E\ket{E}
\ee
and satisfy the conditions
\be\label{norme}
\braket{E}{E^\prime}=\delta (E-E^\prime) \qq{and} \int_0^{+\infty}\ \dd E \ket{E}\bra{E}=\mathbb{1}\ .  
\ee
Following \cite{deAlfaro:1976vlx} we can write $\ket{t}$ as
\be\label{kettE}
\ket{t}= e^{i H t}\ket{t=0}=\int_{0}^{\infty} \dd E\ \frac{\alpha\sqrt{E}}{2} e^{i E t}\ket{E}
\ee
and obtain the overlap between $\ket{E}$ and $\ket{t}$
\be\label{et}
    \braket{t}{E}= \frac{\alpha\sqrt{E}}{2} e^{-i E t}\ .
\ee
 Therefore, the states $\ket{E}$ are similar in spirit to the momentum eigenstates $\ket{\mathbf{p}}$ that one introduces in QFT,  in terms of which the action of a field operator $\phi(\mathbf{x})$ on the vacuum state is 
\be\label{phix}
\phi(\mathbf{x})\ket{0}=\int\frac{\dd^3 p}{(2\pi)^3}\frac{1}{2 E_p} e^{-i \mathbf{p} \cdot\mathbf{x} }\ket{\mathbf{p}}\ , 
\ee
where
\be\label{norme2}
\braket{\mathbf{p}}{\mathbf{p}^\prime}=2 E_\mathbf{p} (2\pi)^3\ \delta^{(3)}(\mathbf{p}-\mathbf{p}^\prime)\ .
\ee
%The analogy is further confirmed when considering
%\be\label{peskin}
%\mel{0}{\phi(\mathbf{x})}{\mathbf{p}}=\bra{0}\int \frac{\dd[3]{p^\prime}}{(2\pi)^3}\ \frac{1}{\sqrt{2 E_{\mathbf{p^\prime}}}}\ \left(a_{\mathbf{p^\prime}}\ e^{i \mathbf{p^\prime}\cdot \mathbf{x}}+ a^\dagger_{\mathbf{p^\prime}}\ e^{-i \mathbf{p^\prime}\cdot \mathbf{x}}\right)\ \sqrt{2 E_{\mathbf{p}}}\ a^\dagger_{\mathbf{p}}\ \ket{0}=e^{i \mathbf{p}\cdot \mathbf{x}}
%\ee
%where \eqref{et} and \eqref{peskin} are exactly the same up to prefactors necessary to satisfy the respective normalisation conditions in \eqref{norme} and \eqref{norme2}.
The analogy between \eqref{kettE} and \eqref{phix} clearly suggests that the state $\ket{t=0}$ plays a role analogous to the inertial vacuum for quantum fields in Minkowski space-time which is the averaging state on which one builds the two-point function. %Such analogy will be corroborated by the considerations in the following section.

\section{``Vacuum" states and horizon temperature}

We have seen that conformal quantum mechanics can be interpreted as a $0+1$ dimensional field theory in which {\it any} generator of conformal transformations of the time axis can be used to define time evolution. Such time evolution can be mapped to motions along $r=\text{const.}$ orbits of time-like conformal Killing vectors in Minkowski space-time. For massless fields in Minkowski space-time one can construct a Fock space using any conformal Killing vector in the domain where the latter is time-like. As in the more familiar case of the Unruh effect \cite{Unruh:1976db}, where boost Killing vectors in the Rindler wedge are used to define positive frequency field modes, Fock spaces constructed using mode decompositions based on different conformal Killing vectors will lead to different Fock spaces and thus to different notions of particles and vacuum states. For the Milne cone and causal diamond in Minkowski space-time these different quantizations were explored in \cite{Higuchi:2017gcd,Wald:2019ygd,Olson:2010jy,Su:2015oys}. These works suggest that, in analogy with the Unruh effect, for both Milne and diamond observers the vacuum state for inertial observers appears as a thermal state.\\

Let us go back to the correspondence between generators of time evolution in conformal quantum mechanics and time-like conformal Killing vectors determining the wordlines of Milne and diamond observers. In light of what we recalled above, we do expect that in conformal quantum mechanics one should be able to identify an inertial ``vacuum" state which is thermally populated by excitations of the Hamiltonian describing the conformal quantum mechanics counterparts of the Milne and diamond time evolutions. This is indeed the case as first shown in \cite{Arzano:2021cjm}. To see this, let us recall that the $\mathfrak{sl}(2,\mathbb{R})$ Lie algebra  \eqref{algebraR} can be realized in terms of two sets of creation and annihilation operators $a^\dagger_L, a^\dagger_R, a_L, a_R$
\begin{equation}\label{algdoppia}
    L_0=\frac{1}{2}\left(a^\dagger_L a_L+a^\dagger_R a_R+1\right)\ , \quad L_+=a^\dagger_L a^\dagger_R \qq{and} L_-=a_L a_R\,.
\end{equation}
This shows that the ground state of the $R$-operator has a {\it bipartite structure}
\begin{equation}
    \ket{n=0}=\ket{0}_L \otimes \ket{0}_R\ ,
\end{equation}
and that the $\ket{t=0}$ state in \eqref{tzeron} can be written as 
\begin{equation}
\label{tvac}
    \ket{t=0}=e^{-a^\dagger_L a^\dagger_R} \ket{0}_L \ket{0}_R=\sum_n\ (-1)^n \ket{n}_L \ket{n}_R = -i 
    \sum_n\ e^{i \pi L_0} \ket{n}_L \ket{n}_R \,.
\end{equation}
From the last equality it is clear that the $\ket{t=0}$ state exhibits a structure similar to that of a {\it thermofield double state} %\cite{Valdivia-Mera:2020nko}
for a harmonic oscillator (see e.g. \cite{Lykken:2020xtx} for a pedagogical review).  Such state can be built by ``doubling" the oscillator's degrees of freedom and is defined by the superposition
\be\label{tfd1}
|TFD\rangle = \frac{1}{Z(\beta)} \sum^{\infty}_{n=0} e^{-\beta E_n/2} |n\rangle_L \otimes |n\rangle_R\,,
\ee
where $Z(\beta) = \sum^\infty_{n=0} e^{-\beta E_n}$ is the partition function at inverse temperature $\beta$. The state \eqref{tfd1} is highly entangled and, tracing over the degrees of freedom of one copy of the system, we obtain a thermal density matrix
\be
Tr_L\{|TFD\rangle \langle TFD|\} =\frac{e^{-\beta H}}{Z(\beta)} 
\ee
at a temperature $T=1/\beta$. The Hamiltonian $H$ is known as {\it modular Hamiltonian}. For a quantum field in Minkowski space-time the inertial vacuum state can be seen as a thermofield double state built on two copies of the Rindler Hilbert space \cite{Valdivia-Mera:2020nko}. Tracing over the degrees of freedom of one copy (i.e. looking at the state from the point of view of Rindler observers whose worldlines are restricted to a space-like wedge) one obtains a thermal state at the Unruh temperature. The modular Hamiltonian in this case is the generator of boosts which can be identified, modulo a factor with dimensions of inverse length related to the magnitude of the acceleration, with the generator of time evolution for Rindler observers.

We see that, setting aside normalization issues which will be the focus of the next section, in conformal quantum mechanics we are dealing with a similar scenario. Indeed we see that tracing over one set of degrees of freedom the reduced density matrix associated to the state \eqref{tvac} has the form of a thermal density matrix for the modular Hamiltonian $-iL_0$ at a temperature $T=1/2\pi$. Since $L_0$ can be identified with the elliptic generator $R$ of the $SO(2)$ compact subgroup of $SL(2,\mathbb{R})$ its ``Wick rotated" counterpart $i L_0$ will generate non compact transformations. Indeed one finds \cite{Arzano:2021cjm} that the generators of the Lie algebra \eqref{so21alg}, besides the identification \eqref{real1}, have two alternative realizations in terms of the generators $H,D$ and $K$ given by
\begin{equation}
\label{algebraS}
    L_0=iS\ ,\qquad L_+=i(D-R)\ ,\qquad L_-=-i(D+R)
\end{equation}
and
\begin{equation}
\label{idstates}
    L_0=iD\ ,\qquad L_+=-i \alpha H\ ,\qquad L_-=-i \frac{K}{\alpha}
\end{equation} 
in which the the modular Hamiltonian $-iL_0$ coincides  with the generators $S$ and $D$. From our discussion in Section 2 we see that, when divided by the constant $\alpha$ with dimensions of length, these two Hamiltonians generate, respectively, translations on diamond and Milne times. 

In the case of the diamond Hamiltonian we notice that the identification \eqref{algebraS} can be obtained by ``Wick rotating" the length parameter $\alpha \rightarrow i \alpha$. Under this map the generator $R$ turns into $iS$ and the wavefunctions \eqref{iptn} into eigenfunctions of the operator $L_0=iS$. Following steps analogous to the ones leading to \eqref{tvac} we find that the state $|t=0\rangle$ has the structure of a thermofield double state for the modular Hamiltonian $S/\alpha$ at a temperature $T=1/(2\pi \alpha)$.

For the Milne Hamiltonian the picture is less straightforward since the generator $D$ is ill defined at $t=0$. One can solve the eigenvalue equation 
\begin{equation}
    (n+1)\, {}_D\langle t|n\rangle  = \mel{t}{i D}{n}=-\left[t\dv{}{t}+1\right]\, {}_D\langle t|n\rangle
\end{equation}
to obtain the eigenstates (see Appendix \ref{cnd})
\begin{equation}\label{tnD}
  {}_D\langle t|n\rangle = \frac{(-1)^n}{2}\ \alpha^{n+2} \sqrt{\frac{\Gamma (2+n)}{n!}}\,  t^{-n-2}\ .
\end{equation}
Since the conformal transformation
 \begin{equation}\label{mappasd}
     t^\prime= \frac{\alpha(t-\alpha)}{\alpha+t}
 \end{equation}
maps the generator $S$ written as a differential operator in terms of the time variable $t'$ into the generator $D$ as a differential operator in the variable $t$, one can obtain the eigenfunctions \eqref{tnD} starting from the eigenfunctions of the $R$ operator \eqref{iptn}, performing the Wick rotation $\alpha \rightarrow i \alpha$ and then the conformal transformation \eqref{mappasd} on the time variable. The states $|t\rangle_D$ can be now written in terms of eigenstates of the $L_0 = i D$ operator as
\begin{equation}\label{tauid}
\begin{split}
    \ket{t}_D&=\frac{1}{2}\left(\frac{\alpha}{t}\right)^2 \sum_{n} (-1)^n \left(\frac{\alpha}{t}\right)^n \sqrt{\frac{\Gamma (2+n)}{n!}} \ket{n}=\frac{1}{2}\left(\frac{\alpha}{t}\right)^2 e^{- \frac{\alpha}{t}L_+}\ket{n=0}\ .    
\end{split}
\end{equation}
 We see that now the state $t=\alpha$ exhibits the structure of a thermofield double state for the modular Hamiltonian $D/\alpha$ at the temperature $T=1/(2\pi \alpha)$. The point $t=\alpha$ is the image of the origin under the conformal mapping \eqref{mappasd} and it corresponds to the origin of the conformal time $\tau$ variable defined by $t = \alpha\ e^{ \frac{\tau}{\alpha}}$.

The state $\ket{t=0}$, as evidenced by eq. \eqref{tvac}, is an entangled state with respect to the bi-partition of the Hilbert space in terms of $L$ and $R$ degrees of freedom. In analogy with the case of the inertial vacuum written as a thermofield double state over the excitations of the left and right Rindler modes (the two complementary domains of the evolution of the boost modular Hamiltonian), we can think of the two sets of degrees of freedom in \eqref{tvac} as belonging to the domain of diamond and Milne time evolution and their complements (the restriction to $r=0$ wordlines of the entanglement considered in \cite{Higuchi:2017gcd} and \cite{Olson:2010jy}).

We can quantify this entanglement by calculating the Von Neumann entropy of the reduced density matrix obtained by tracing over one set of degrees of freedom in the density matrix associated to the inertial vacuum $\ket{t=0}$. Such entanglement entropy can be seen as the $0+1$-dimensional analogue of the entanglement entropy a quantum field across space-time regions. Unlike its higher dimensional counterparts, the simple structure of the state $\ket{t=0}$ makes it rather straightforward to calculate the entanglement entropy associated to the diamond and Milne time domains.

\section{Entanglement entropy}
In order to derive the entanglement entropy associated to the partition of the $\ket{t=0}$ state we first notice that such state is non-normalizable. This is to be expected given the correspondence between the inner product $\braket{t_1}{t_2}$ and the restriction of the two-point function of a massless field to the $r=0$ worldline since it reflects the UV divergence of the latter for coincident points.

We can regularize the state $\ket{t=0}$ via an infinitesimal translation in imaginary time. We consider a state at time $t=i\epsilon$ 
\begin{equation}
\label{eq2}
\ket{t=i\epsilon}=\left(\frac{\frac{\alpha-\epsilon}{\alpha+\epsilon}+1}{2}\right)^{2}\ e^{-\frac{\alpha-\epsilon}{\alpha+\epsilon}\ L_+}\ket{n=0}
\end{equation}
where $\epsilon$ can be interpreted as a short-distance cut-off scale. We have 
\begin{equation}
    \ket{t=i\epsilon}=\left(\frac{\alpha}{\alpha+\epsilon}\right)^{2} \sum_{n=0}^{\infty}(-1)^n\left(\frac{\alpha-\epsilon}{\alpha+\epsilon}\right)^n \ket{n}_L\ket{n}_R
\end{equation}
so that
\begin{equation}
    \bra{t=i\epsilon}\ket{t=i\epsilon}=\frac{\alpha }{4\epsilon}\left(\frac{\alpha}{\alpha+\epsilon}\right)^{2}\equiv \frac{1}{\mathcal{N}^2}\ .
\end{equation}

 In order to derive the reduced density matrix we normalize the $\ket{t=i\epsilon}$ and introduce a state $\ket{\delta}$ with unitary norm
\begin{equation}
    \ket{\delta} \equiv \mathcal{N} \ket{i\epsilon}\ .
\end{equation}
Let us now consider the density matrix
\begin{equation}
\label{rhoLR}
    \rho_{RL}=\ket{\delta}\bra{\delta}
\end{equation}
and compute the reduce density matrix $\rho_L$ explicitly by tracing over the R degrees of freedom
\begin{equation}
\label{rhoL}
    \rho_L=\Tr_R{\rho_{LR}}=\mathcal{N}^2 \left(\frac{\alpha}{\alpha+\epsilon}\right)^4 \sum_{n=0}^\infty \left(\frac{\alpha-\epsilon}{\alpha+\epsilon}\right)^{2n} \ket{n}_L\bra{n}_L \ .
\end{equation}
The Von Neumann entropy of the reduced density matrix is then given by
\begin{equation}
    S=-\Tr{\rho_L\log\rho_L}=-\frac{(\alpha-\epsilon )^2 \log \left(\frac{(\alpha-\epsilon )^2}{(\alpha+\epsilon )^2}\right)}{4 \alpha \epsilon }-\log \left(\frac{4 \alpha \epsilon }{(\alpha+\epsilon )^2}\right)\,,
\end{equation}
which considering the limit $\epsilon \rightarrow 0$ leads to
\begin{equation}
\label{nostro}
    S=\log \left(\frac{\alpha}{\epsilon}\right)+\text{const}+\order{\epsilon^2}\, .
\end{equation}
We see that the result obtained exhibits a logarithmic divergence when the UV cut-off scale $\epsilon$ is sent to zero. Let us recall that in a d-dimensional free quantum field theory the entanglement entropy associated to a spatial region $\mathcal{A}$ is UV divergent with a leading divergent term proportional to
\be
\frac{\text{Area}(\partial \mathcal{A})}{\epsilon^{d-2}}\ ,
\ee
where $\text{Area}(\partial \mathcal{A})$ is the area of the boundary of the region (entangling surface) and $\epsilon$ is a UV cut-off \cite{Rangamani:2016dms}.
%and has the general expressions \cite{Rangamani:2016dms}
%\begin{align}
%    S_\mathcal{A}&=g_{d-2} \left(\frac{\alpha}{\epsilon}\right)^{d-2}+g_{d-4} \left(\frac{\alpha}{\epsilon}\right)^{d-4}+\dots +g_1 \left(\frac{\alpha}{\epsilon}\right)+ (-1)^{\frac{d-1}{2}} g_0+ \mathcal{O}(\epsilon)\qquad  &\qq{d odd}\\\label{entropycft}
 %   S_\mathcal{A}&= g_{d-2} \left(\frac{\alpha}{\epsilon}\right)^{d-2}+g_{d-4} \left(\frac{\alpha}{\epsilon}\right)^{d-4}+\dots+ (-1)^{\frac{d-2}{2}} g_0\log\left(\frac{\alpha}{\epsilon}\right)+ \mathcal{O}(\epsilon^0)  \qquad &\qq{d even}
%\end{align}
%with $d$ being the dimension of the spacetime, $g_i$ with $i>0$ depend on the regularization prescription, $\alpha$ is proportional to the size of the region $\mathcal{A}$ and $\epsilon$ is again an UV regulator. Particularly important is the coefficient $g_0$ which carries non-trivial and universal information. 
Two-dimensional conformal theories $\mathrm{CFT_2}$ are a special case since they predict a logarithmic divergence and therefore the entanglement entropy fails to follow an area law. For example the entanglement entropy between a shorter line-segment with length $\alpha$ and a longer one with length $L$ containing it in the limit  $\frac{\alpha}{L}\ll 1$ reads \cite{Calabrese:2004eu,Saravani:2013nwa,Rangamani:2016dms}
\begin{equation}
\label{rid}
    S=\frac{c}{3}\log{\frac{\alpha}{\epsilon}}+\text{const}\ ,
\end{equation}
where $c$ is the central charge of the CFT which is equal to $1$ in a quantum field theory of a massless bosonic field.
This peculiar behaviour can be understood heuristically by arguing that the logarithm arises as a limiting case of a power law divergence and it is consistent with the entangling surface comprising of a set of disconnected points. 

We observe that the analytical behaviour of our result \eqref{nostro} for the entanglement entropy in $\mathrm{CFT_1}$ is the same as in $\mathrm{CFT_2}$ and in particular it shows the same logarithmic divergence. This is in line with the point-like nature of the entangling surface.

\section{Conclusions}
Conformal quantum mechanics is a simple one-dimensional model which, as we have seen, possesses enough structure to mimic the non-trivial vacuum structure of higher dimensional free quantum field theories. Such non-trivial structure is due to the existence of excitations in the theory which are associated to Hamiltonians whose orbits posses a boundary, the one dimensional counterparts of horizons in higher dimensions. As in quantum field theories in higher dimensions we have seen that one can associate temperatures to these horizons. Moreover, we evaluated the entanglement entropy associated to the bipartite decomposition of the states on which one builds the two-point function of the theory into modes of Hamiltonians whose orbits do not cover the entire time domain. The possibility of entanglement between time domains in Minkowski space-time has been considered before \cite{Olson:2010jy,Higuchi:2017gcd}. Here we provide an explicit calculation of entanglement entropy in conformal quantum mechanics for partitions of states in terms of modular Hamiltonians which define evolution in time domains with boundaries. These domains can be embedded in Minkowski space-time as wordlines of diamond and Milne observers at the origin and, thus, one could interpret our result as quantifying a {\it worldline entanglement entropy} (see \cite{Anninos:2011af,Nakayama:2011qh} for a similar interpretation of conformal quantum mechanics as worldline quantum mechanics for static patch observers in de Sitter space-time) for observers who cannot have access to the past beyond a certain point (the initial ``singularity" for Milne observers) or to the future and the past beyond two points (the future and past tips of the diamond for diamond observers). The worldline boundaries are point-like and the entanglement entropy exhibits a logarithmic divergence similar to that found in two-dimensional CFTs, where the boundaries of the spatial region considered are also point-like. It is tempting to speculate that such worldline entropy could play a role as a tool for facilitating the calculation of entanglement entropy for conformally invariant quantum field theories in higher space-time dimensions. We leave this task for future investigations.

\section*{Acknowledgements}
We acknowledge support from the INFN Iniziativa Specifica  QUAGRAP.  This research was carried out in the frame of Programme STAR Plus, financially supported by the University of Napoli Federico II and Compagnia di San Paolo. This work also falls within the scopes of the  European Union COST Action CA18108 {\it Quantum gravity phenomenology in the multi-messenger approach}.

\begin{appendices}
 \renewcommand{\theequation}{\thesection.\arabic{equation}}

\section{Determining the coefficients $c_n$}\label{cn}
In order to find the coefficients $c_n$ we act with the ladder operators. We start with the action of $L_+$
\begin{align}
    \sqrt{(n+1)(n+2)}\ip{t}{n+1}&=\mel{t}{L_+}{n}=\left[\left(i\frac{t}{\alpha}-1\right)+\left(i\frac{t^2}{2\alpha}-i\frac{\alpha}{2}-t\right)\dv{}{t}\right]\ip{t}{n}
\end{align}
which gives
\begin{equation}
    \frac{c_n}{c_{n+1}}=\frac{(n+1) }{\sqrt{(n+1) (n+2)}}
\end{equation}
while acting with $L_-$
\begin{align}
    \sqrt{n(n+1)}\ip{t}{n-1}&=\mel{t}{L_-}{n}=\left[\left(i\frac{t}{\alpha}+1\right)+\left(i\frac{t^2}{2\alpha}-i\frac{\alpha}{2}+t\right)\dv{}{t}\right]\ip{t}{n}
\end{align}
we arrive at
\begin{equation}
    \frac{c_{n-1}}{c_{n}}=\frac{n}{\sqrt{n (n+1)}}\ ,
\end{equation}
hence we conclude that
\begin{equation}
    c_n=\sqrt{\frac{\Gamma(2+n)}{\Gamma(n+1)}}=\sqrt{\frac{\Gamma(2+n)}{n!}}\ .
\end{equation}

\section{Eigenstates of the $iD$ generator}\label{cnd}

The differential equation 
\begin{equation}
    (n+1)\ip{t}{n}=\mel{t}{i D}{n}=-\left[t\dv{}{t}+1\right]\ip{t}{n}
\end{equation}
admits solutions
\begin{equation}
  \ip{t}{n}= c_n\  t^{-n-2}\ .
\end{equation}
To determine the coefficients $c_n$ we act with $L_+$ on these functions
\begin{align}
    \sqrt{(n+1)(n+2)}\ip{t}{n+1}&=\mel{t}{L_+}{n}= \alpha \dv{}{t}\ip{t}{n}
\end{align}
and obtain
\begin{equation}
    \sqrt{(n+1)(n+2)}\ c_{n+1}=- \alpha\ c_n (2+n)\ .
\end{equation}
The action with $L_-$ gives
\begin{align}
    \sqrt{n(n+1)}\ip{t}{n-1}&=\mel{t}{L_-}{n}=\frac{1}{\alpha}\left[t^2 \dv{}{t}+2t\right]\ip{t}{n}
\end{align}
whose solution is
\begin{equation}
    - \alpha \sqrt{n(n+1)} c_{n-1}=n\  c_n\ .
\end{equation}
By combining the two results obtained we arrive at the following expression for the coefficients $c_n$
\begin{equation}
    c_n= \frac{(-1)^n}{2}\ \alpha^{n+2} \sqrt{\frac{\Gamma (2+n)}{n!}}\ .
\end{equation}

\end{appendices}

\bibliography{bibliography}

\end{document}